\documentclass[a4paper]{jpconf}
\usepackage{graphicx}
\usepackage{hyperref}
\usepackage{float}
\usepackage{booktabs}
\begin{document}
\title{Towards energy discretization for muon scattering tomography in GEANT4 simulations: A discrete probabilistic approach}

\author{A. Ilker Topuz$^{\rm 1,2}$, Madis Kiisk$^{\rm 1,3}$}

\address{$^{1}$ Institute of Physics, University of Tartu, W. Ostwaldi 1, 50411, Tartu, Estonia}
\address{$^{2}$ Centre for Cosmology, Particle Physics and Phenomenology, Universit\'e catholique de Louvain, Chemin du Cyclotron 2, B-1348 Louvain-la-Neuve, Belgium}
\address{$^{3}$ GScan OU, Maealuse 2/1, 12618 Tallinn, Estonia}

\ead{ahmet.ilker.topuz@ut.ee, ahmet.topuz@uclouvain.be}

\begin{abstract}
In this study, by attempting to eliminate the disadvantageous complexity of the existing particle generators, we present a discrete probabilistic scheme adapted for the discrete energy spectra in the GEANT4 simulations. In our multi-binned approach, we initially compute the discrete probabilities for each discrete energy bin, the number of which is flexible depending on the computational goal, and we solely satisfy the imperative condition that requires the sum of the discrete probabilities to be the unity. Regarding the implementation in the GEANT4 code, we construct a one-dimensional probability grid that consists of sub-cells equaling the number of the energy bin, and each cell represents the discrete probability of each energy bin by fulfilling the unity condition. Through uniformly generating random numbers between 0 and 1, we assign the discrete energy in accordance with the associated generated random number that corresponds to a specific cell in the probability grid. This probabilistic methodology does not only permits us to discretize the continuous energy spectra based on the Monte Carlo generators, but it also gives a unique access to utilize the experimental energy spectra measured at the distinct particle flux values. Ergo, we initially perform our simulations by discretizing the muon energy spectrum acquired via the CRY generator over the energy interval between 0 and 8 GeV along with the measurements from the BESS spectrometer and we determine the average scattering angle, the root-mean-square of the scattering angle, and the number of the muon absorption by using a series of slabs consisting of aluminum, copper, iron, lead, and uranium. Eventually, we express a computational strategy in the GEANT4 simulations that grants us the ability to verify as well as to modify the energy spectrum depending on the nature of the information source in addition to the exceptional tracking speed. 
\end{abstract}
\section{Introduction}
The emerging applications of cosmic ray muon tomography~\cite{pesente2009first, procureur2018muon, bonechi2020atmospheric} lead to a significant rise in the utilization of the cosmic particle generators, e.g. CRY~\cite{hagmann2007cosmic}, CORSIKA~\cite{heck1998corsika}, or CMSCGEN~\cite{biallass2009parametrization}, where the fundamental parameters such as the energy spectrum and the angular distribution associated with the generated muons are represented in the continuous forms routinely governed by the probability density functions over the corresponding vast intervals. Despite this perceptible increase in the diversity of the muon generators, the common difficulties in the hands-on applications that might be relatively exemplified as perplexing coupling with the Monte Carlo codes, unnecessarily broad and occasionally unmodifiable parametric intervals for the specific applications, extensive execution times, and complications in the particle tracking partly remain steady. Contrary to the continuous mode, the discretized energy spectra, i.e. multi-group energy approximations to put it another way, have been ubiquitously employed in the neighboring fields such as nuclear engineering~\cite{Dermott, nakagawa1984morse} and medical physics~\cite{rivard2001discretized, boman2004modelling} under the umbrella of the non-analogue Monte Carlo simulations~\cite{shultis2011mcnp} on and on. Along with the discretization schemes based on the theoretical assumptions, a number of notable empirical studies founded on the advanced particle detectors such as the BESS spectrometer~\cite{Haino2004measurements} represent the experimental energy spectra in the discrete format. While MNCP6~\cite{goorley2012initial} includes the necessary algorithms to utilize the discrete energy distributions in the black box paradigm, the general particle source (GPS) in GEANT4~\cite{agostinelli2003geant4} is the existing pre-configured module that provides the opportunity of this discrete approach through a macro file without detailing the algebraic/algorithmic phase. 

In the present study, by aiming at obtaining a fast, verifiable, and modifiable muon source in the GEANT4 simulations, we initially exhibit our procedure that is based on the multi-binned approximation of the CRY muon energy spectrum. In the latter instance, we incorporate our discrete spectra in the GEANT4 code by using a one-dimensional probability grid that operates under G4ParticleGun. Following this step, we furthermore gain the capability of utilizing the existing experimental spectra. Then, we test our methodology over our tomographic setup consisting of plastic scintillators fabricated from polyvinyl toluene with the dimensions of 100$\times$0.4$\times$100 $\rm cm^{3}$ and we determine the characteristic parameters such the average scattering angle, the root-mean-square of the scattering angle, and the number of the muon absorption by using a set of slabs composed of aluminum, copper, iron, lead, and uranium with the dimensions of 40$\times$10$\times$40 $\rm cm^{3}$. Finally, we contrast our simulation outcomes by means of both the CRY discrete spectrum and the BESS muon spectrum. This study is outlined as follows. In section~\ref{Discreteenerprob}, we compute the discrete probabilities for the corresponding discrete energies, and section~\ref{Implement} describes the implementation in the GEANT4 code. While we express our characteristic parameters as well as our simulation setup in section~\ref{Characsim}, we expose our simulation outcomes in section~\ref{results}. Finally, our conclusions are stated in section~\ref{Conclusion}.
\section{Discrete energies and discrete probabilities}
\label{Discreteenerprob}
\begin{figure}[H]
\begin{center}
\includegraphics[width=7.9cm]{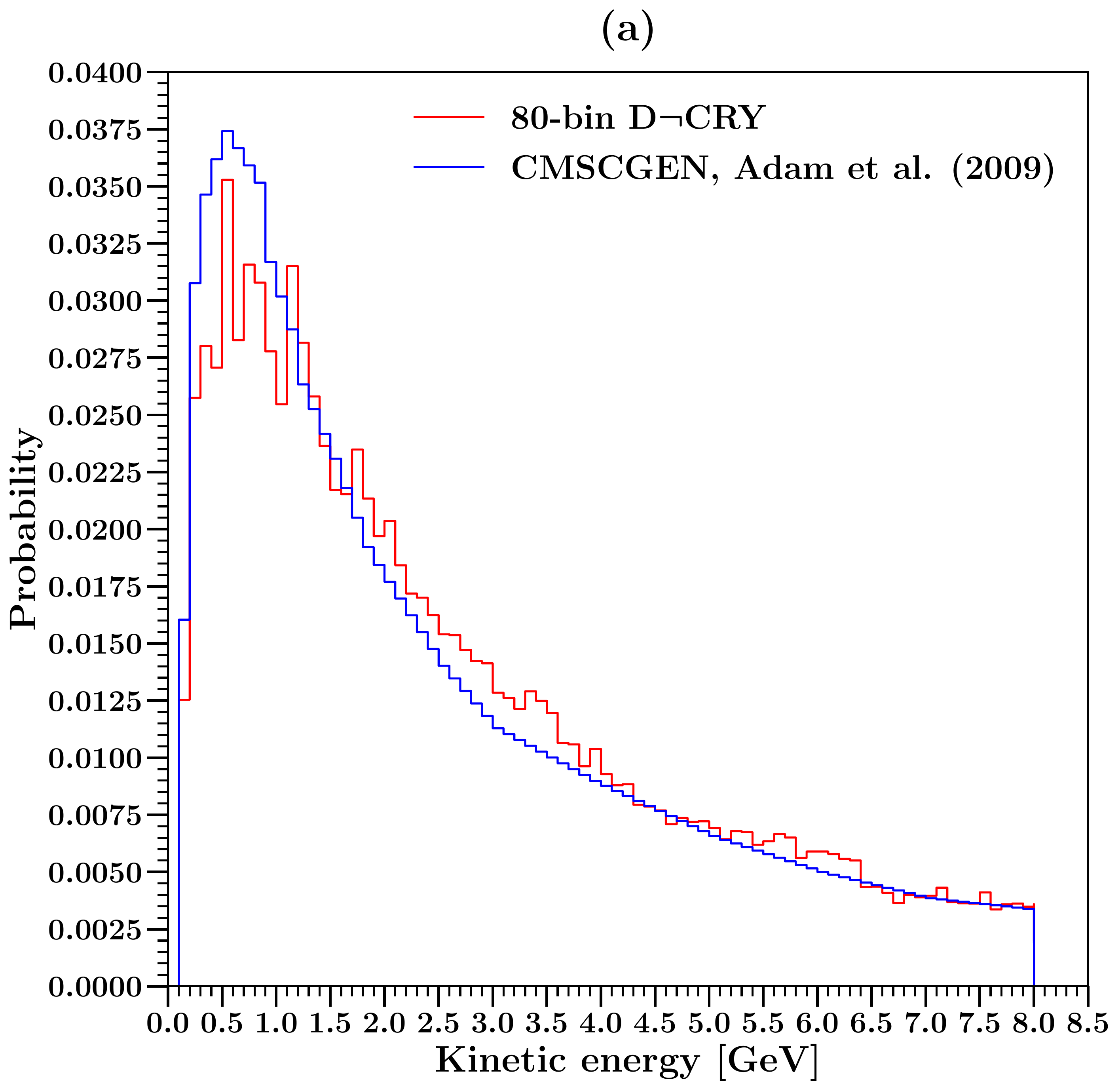}
\includegraphics[width=7.9cm]{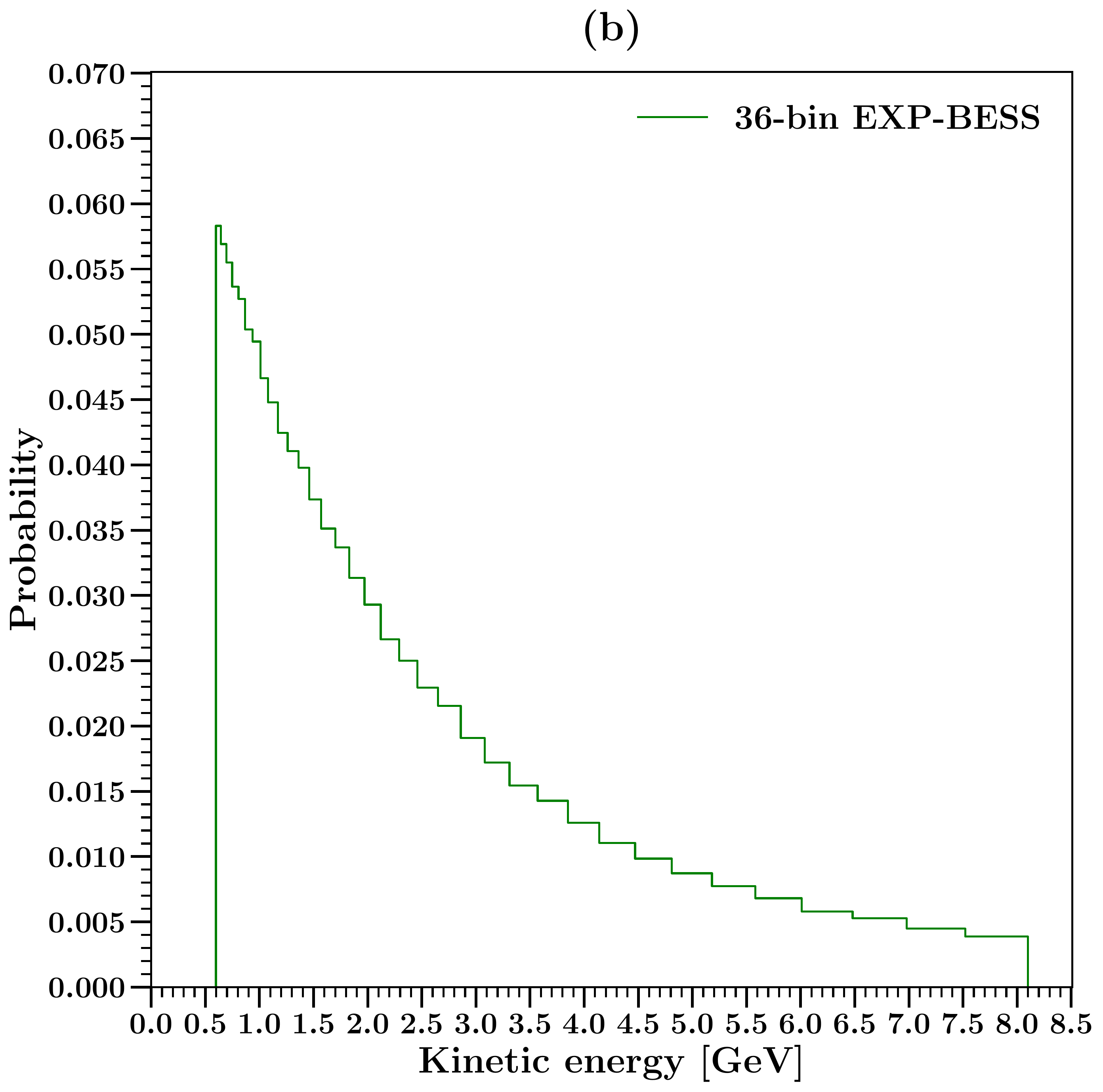}
\vspace{-2em}
\caption{Comparison between discrete energy histograms (a) 80-bin D$\neg$CRY compared to CMSCGEN and (b) 36-bin EXP-BESS.}
\label{energyhisto}
\end{center}
\vskip -0.5cm
\end{figure}

At the outset, we strive for the energy discretization based on the extracted energy list from the CRY muon generator~\cite{hagmann2007cosmic} between 0 and 8 GeV. To achieve this aim, we first set out our constant bin length that is selected as 0.1 GeV. Thus, the number of non-zero energy bins is evidently 80 as written in 
\begin{equation}
\#_{\rm Bins}=\frac{E_{\rm Max} - E_{\rm Min}}{L_{\rm Bins}}=\frac{8.0-0}{0.1}=80
\end{equation}
Then, in agreement with the energy dataset acquired through the CRY generator, the number of the counts in the specific energy bin denoted by $E_{i}$ in GeV is computed by incorporating any $E_{x}\in(E_{i-1}, E_{i}]$ under the condition of $m_{0}=0$ for $E_{0}=0$ as described in
\begin{equation}
m_{i}=\sum\limits_{k=1}1\mbox{	if $E_{i-1}<E_{x}\leq E_{i}$	for i={1, 2, 3,.., 80}}
\label{Count}
\end{equation}
Basically, Eq.~(\ref{Count}) might be performed by using the existing tools in Python, e.g. NumPy. While this operation is an approximation by definition, we do not significantly neglect any fundamental information in the case of the fine energy bins as accomplished 
in the present study, which also implies that the simulation outcomes by means of the finely binned histogram do not yield consequential differences in comparison with the continuous energy spectra. Whereas the count numbers in the particular energy bins already provides an opportunity to employ the discrete energy distributions in the wake of renormalization, we favor to determine the discrete probabilities that serve to constitute a probability grid; hence, we first calculate the total count over 80 energy bins as shown in 
\begin{equation}
\sum\limits_{i=0}^{\#_{\rm Bins}} m_{i}=\sum\limits_{i=0}^{80} m_{i}
\end{equation}
In the end, the discrete probability, i.e. the discrete normalized frequency to rephrase it, at a given energy bin indicated by $E_{i}$ is the ratio between the specific count in $E_{i}$ and the total counts over 80 bins by satisfying the unity condition as noted in
\begin{equation}
p_{i}=\frac{m_{i}}{\sum\limits_{i=0}^{80}m_{i}}~~~{\rm with}~~~\sum\limits_{i=0}^{80}p_{i}=1
\end{equation}
Finally, we tabulate the discrete probabilities along with the discrete muon energies obtained via the CRY muon source in Table~\ref{CRY} by entitling D$\neg$CRY. As indicated in Table~\ref{CRY}, we view that the highest discrete probability appertains to the energy bin of 0.5 GeV, i.e. the mode of our discrete energy distribution, which is reduced approximately by one order of magnitude at 8 GeV, and the trend observed in D$\neg$CRY resembles to a log-normal distribution with a positive skewness. 
\vskip -0.5cm
\begin{table}[H]
\begin{center}
\caption{D$\neg$CRY discrete probabilities between 0 and 8 GeV.}
\begin{tabular}{cc}
\toprule
\toprule
$E_{i}$ [GeV] & $p_{i}$\\
\midrule
0.0&0.000000\\
0.1&0.012536\\
0.2&0.025745\\
0.3&0.028020\\
0.4&0.027066\\
0.5&0.035285\\
0.6&0.028265\\
0.7&0.031579\\
0.8&0.030784\\
0.9&0.027776\\
1.0&0.025464\\
1.1&0.031506\\
1.2&0.028155\\
1.3&0.025807\\
1.4&0.023642\\
1.5&0.021709\\
1.6&0.021526\\
1.7&0.023483\\
1.8&0.021342\\
1.9&0.019691\\
2.0&0.020364\\
2.1&0.018419\\
2.2&0.017184\\
2.3&0.017001\\
2.4&0.016242\\
2.5&0.015398\\
2.6&0.015362\\
\bottomrule
\bottomrule
\end{tabular}
\begin{tabular}{cc}
\toprule
\toprule
$E_{i}$ [GeV] & $p_{i}$\\
\midrule
2.7&0.014713\\
2.8&0.014224\\
2.9&0.014126\\
3.0&0.012842\\
3.1&0.012610\\
3.2&0.012133\\
3.3&0.012903\\
3.4&0.012487\\
3.5&0.011962\\
3.6&0.010641\\
3.7&0.010579\\
3.8&0.009626\\
3.9&0.010384\\
4.0&0.009283\\
4.1&0.008794\\
4.2&0.008843\\
4.3&0.007938\\
4.4&0.007864\\
4.5&0.007693\\
4.6&0.007094\\
4.7&0.007363\\
4.8&0.007192\\
4.9&0.007216\\
5.0&0.006923\\
5.1&0.006433\\
5.2&0.006788\\
5.3&0.006739\\
\bottomrule
\bottomrule
\end{tabular}
\begin{tabular}{cc}
\toprule
\toprule
$E_{i}$ [GeV] & $p_{i}$\\
\midrule
5.4&0.006189\\
5.5&0.006348\\
5.6&0.006653\\
5.7&0.006507\\
5.8&0.005614\\
5.9&0.005895\\
6.0&0.005895\\
6.1&0.005785\\
6.2&0.005577\\
6.3&0.005504\\
6.4&0.004342\\
6.5&0.004354\\
6.6&0.004085\\
6.7&0.003645\\
6.8&0.003999\\
6.9&0.003889\\
7.0&0.003963\\
7.1&0.004317\\
7.2&0.003681\\
7.3&0.003632\\
7.4&0.003620\\
7.5&0.004109\\
7.6&0.003363\\
7.7&0.003584\\
7.8&0.003620\\
7.9&0.003486\\
8.0&0.003596\\
\bottomrule
\bottomrule
\end{tabular}
\label{CRY}
\end{center}
\end{table}
In view of the fact that we aim at carrying out a multi-group approach, we are furthermore capable of utilizing the empirical muon energy spectrum as in the case of the BESS spectrometer~\cite{Haino2004measurements} where the discrete energies are measured at the distinct particle fluxes. We adopt the first 36 non-zero energy bins between 0.598 and 8.1 GeV and their corresponding separate fluxes by labeling as EXP-BESS, and the discrete probabilities are determined by the quotient between the particular flux value at a specific energy and the total flux as follows
\begin{equation}
p_{i}=\frac{\phi_{i}}{\sum\limits_{i=0}^{36}\phi_{i}}~~~{\rm with}~~~\sum\limits_{i=0}^{36}p_{i}=1
\label{BESSprob}
\end{equation}
Since the energy list is already discretized, we exert neither a further approximation nor a simplifying assumption in Eq.~(\ref{BESSprob}). The experimental energy values obtained through the BESS spectrometer as well as the computed discrete probabilities are listed in Table~\ref{BESSlist}.
\vskip -0.5cm
\begin{table}[H]
\setlength{\tabcolsep}{2em}
\begin{center}
\caption{Discrete EXP-BESS probabilities between 0 and 8.1 GeV.}
\begin{tabular}{*2c}
\toprule
\toprule
$E_{i}$ [GeV] & $p_{i}$\\
\midrule
0.000&0.000000\\
0.598&0.058310\\
0.644&0.056910\\
0.694&0.055511\\
0.748&0.053645\\
0.806&0.052712\\
0.868&0.050379\\
0.936&0.049447\\
1.010&0.046648\\
1.080&0.044782\\
1.170&0.042449\\
1.260&0.041050\\
1.360&0.039790\\
1.460&0.037365\\
1.570&0.035126\\
1.700&0.033680\\
1.830&0.031347\\
1.970&0.029295\\
2.120&0.026636\\
\bottomrule
\bottomrule
\end{tabular}
\begin{tabular}{*2c}
\toprule
\toprule
$E_{i}$ [GeV] & $p_{i}$\\
\midrule
2.290&0.025003\\
2.460&0.022951\\
2.650&0.021551\\
2.860&0.019079\\
3.080&0.017213\\
3.310&0.015440\\
3.570&0.014274\\
3.850&0.012595\\
4.140&0.011055\\
4.470&0.009843\\
4.810&0.008723\\
5.180&0.007744\\
5.580&0.006811\\
6.010&0.005784\\
6.480&0.005271\\
6.980&0.004478\\
7.520&0.003872\\
8.100&0.003233\\
\\
\bottomrule
\bottomrule
\end{tabular}
\label{BESSlist}
\end{center}
\end{table}
As can be noticed from Table~\ref{BESSlist}, the EXP-BESS muon spectrum exhibits an exponentially decreasing trend where the lower bound is associated with the highest discrete probability, whereas the upper bound is identified with the minimum discrete normalized frequency. In order to qualitatively show the variation of both D$\neg$CRY and EXP-BESS, Fig.~\ref{energyhisto} displays the energy histograms where (a) indicates D$\neg$CRY in comparison with the discrete CMSCGEN data reproduced from another study~\cite{adam2009performance} that is dedicated to the CMS strip tracker, while (b) points out EXP-BESS. From Fig.~\ref{energyhisto}(a), we experience that our D$\neg$CRY tends to show strong similarities when compared to the existing qualitative data of the CMSCGEN muon generator, thereby relatively authenticating our discretized energy spectrum hinged on the CRY muon generator.
\section{Implementation via G4ParticleGun in GEANT4}
\label{Implement}
\begin{figure}[H]
\begin{center}
\includegraphics[width=10.5cm]{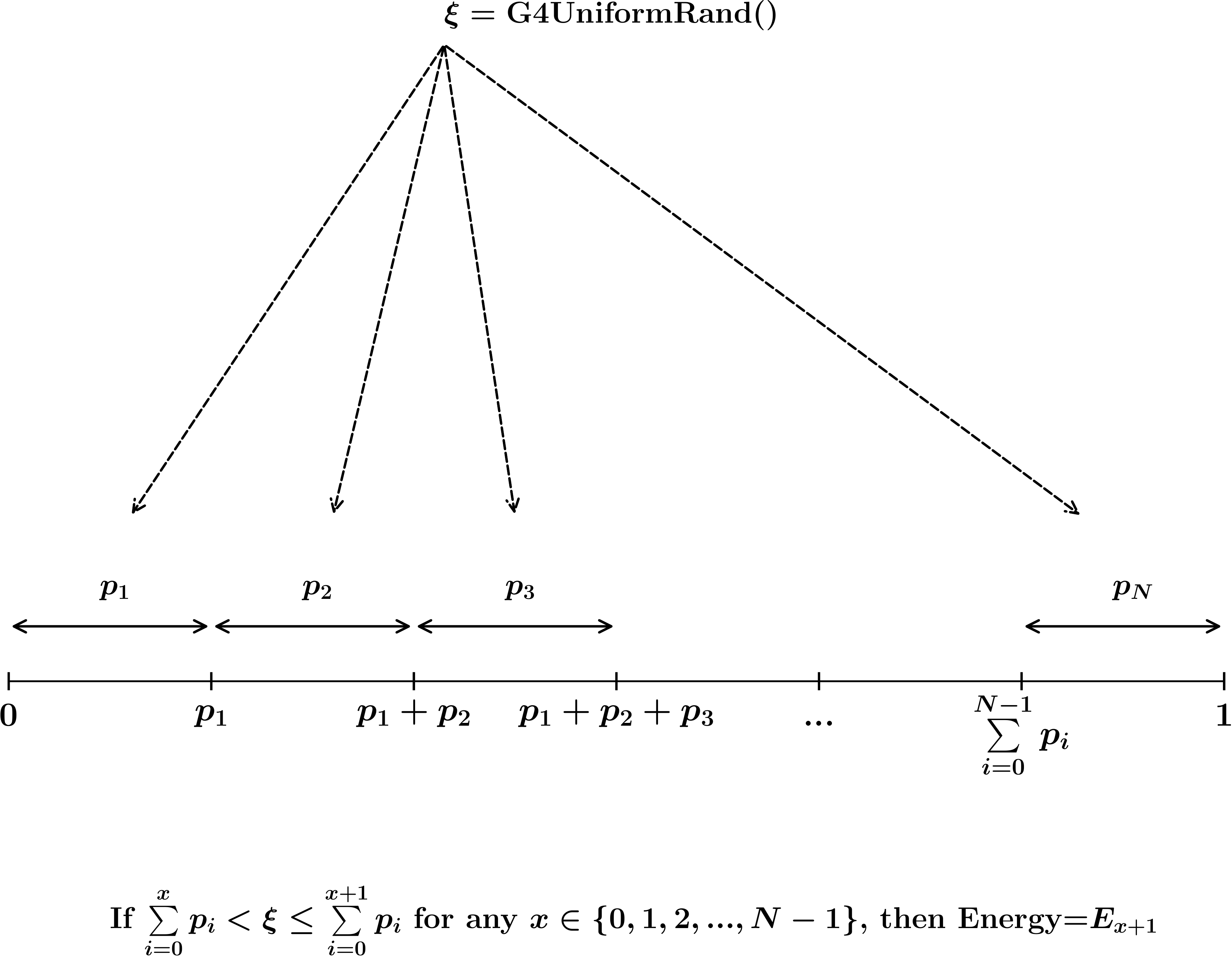}
\caption{Scheme of probability grid implemented via G4ParticleGun in GEANT4.}
\label{generation}
\end{center}
\end{figure}

\begin{figure}[H]
\begin{center}
\includegraphics[width=7.9cm]{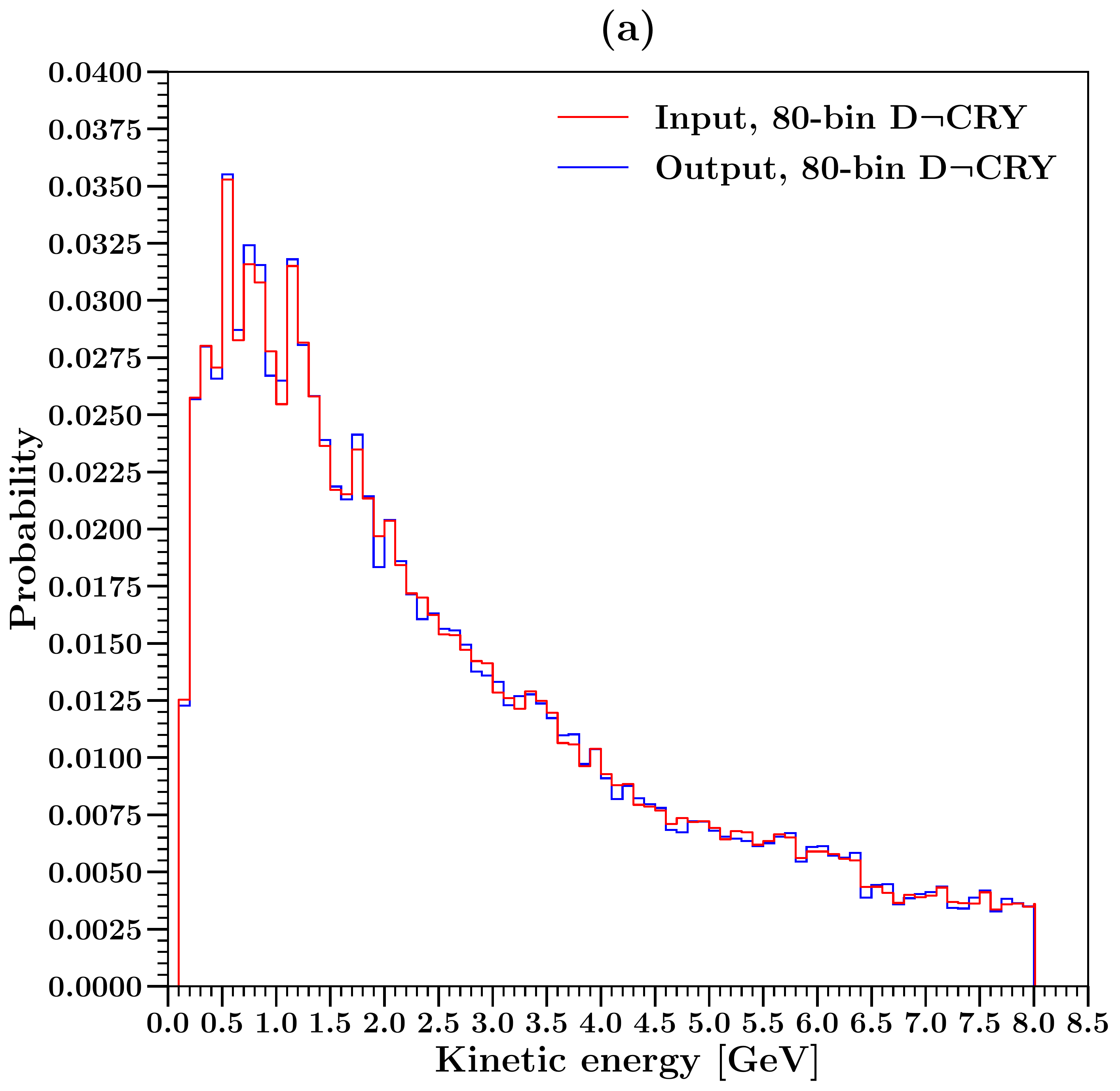}
\includegraphics[width=7.9cm]{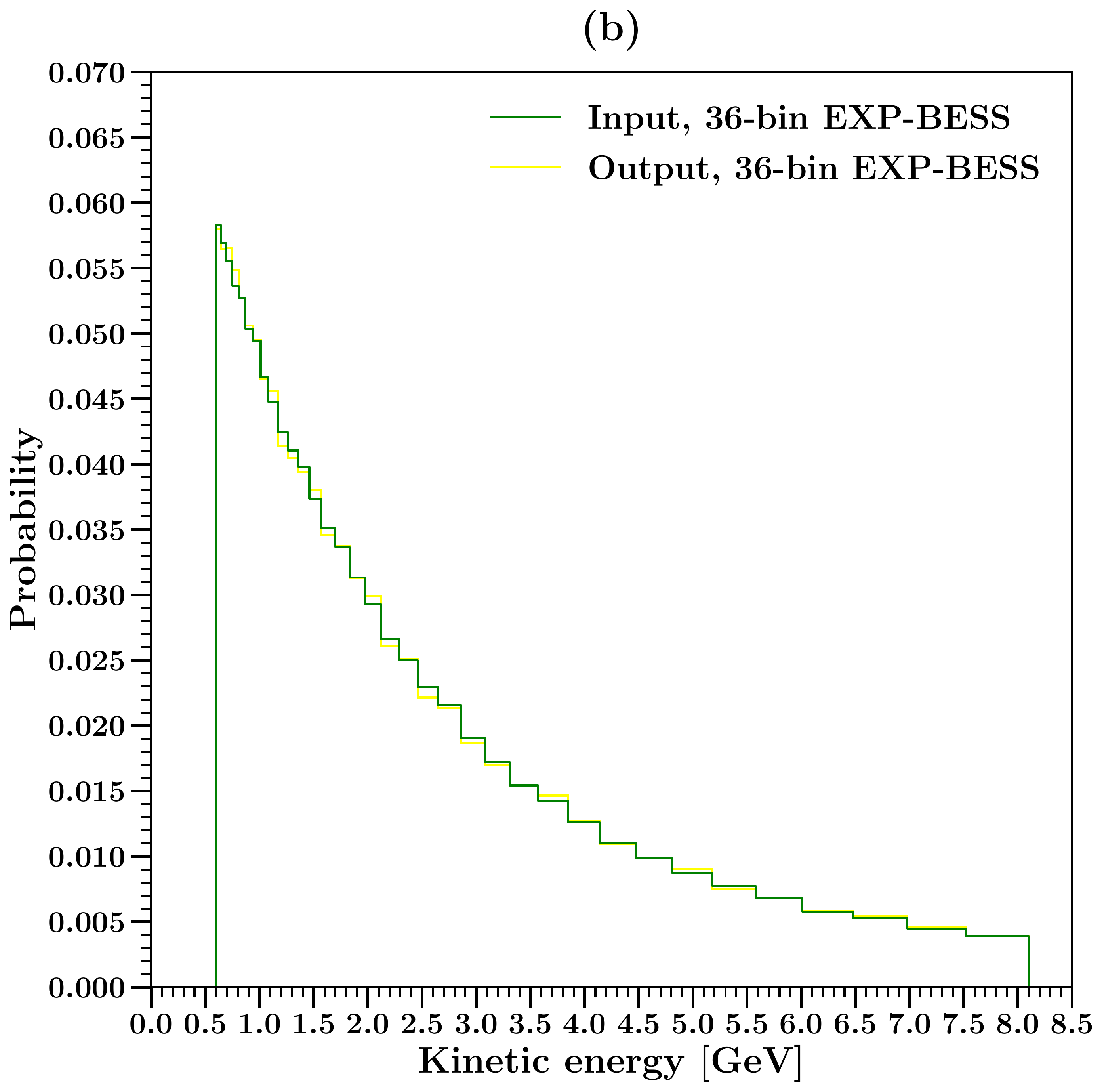}
\vspace{-2em}
\caption{Contrast between the input dataset before processing with the probability grid and the output values through the activation of our probability grid (a) 80-bin D$\neg$CRY and (b) 36-bin EXP-BESS.}
\label{gridper}
\end{center}
\end{figure}

On account of attaining the discrete energy spectra, we subsequently integrate our strategy to inject the incoming muons by means of G4ParticleGun. By reminding the unity condition, we build a grid by adding up the discrete probabilities, the inverval of which starts with 0 and ends in 1 as illustrated in Fig.~\ref{generation}. Thus, each cell in this grid, i.e. the difference between two points on the probability grid, specifies a discrete probability. Then, we generate a random number denoted by $\xi$ between 0 and 1 by using the pre-defined uniform number generator called G4UniformRand(). Finally, we scan this random number on the probability grid by checking the difference between the grid points and we assign the particular discrete energy when the random number matches with the associated cell.
Intending to reveal out the performance of our procedure, the energy histrograms before and after the probability grid are put on view in Fig.~\ref{gridper} where the term called input indicates the discrete energy values as well as the discrete probabilities that are stored in an array before the activation of the probability grid, whereas the notion named output hints the outcome of the probability grid by processing the input information, which means the kinetic energies allocated to the generated muons by means of the probability grid.
\section{Characteristic parameters and simulation properties}
\label{Characsim}
\begin{figure}[H]
\begin{center}
\includegraphics[width=9cm]{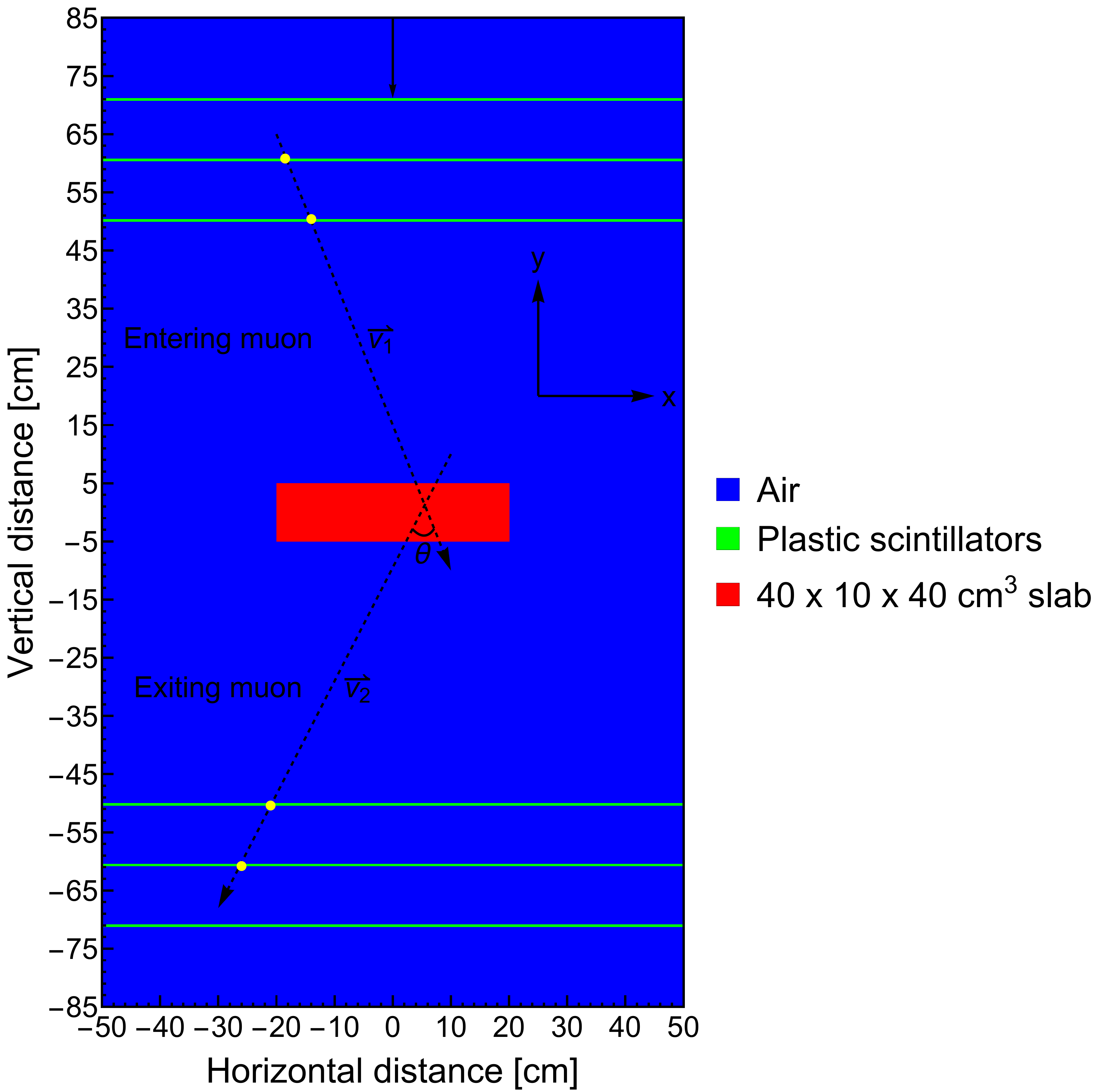}
\caption{Delineation of scattering angle within the present tomographic configuration.}
\label{scatteringangle}
\end{center}
\end{figure}

For the purpose of appraisal, we express a number of characteristic parameters consorted with the muon tomography. By defining two vectors, we bring forth the scattering angle denoted by $\theta$ as depicted in Fig.~\ref{scatteringangle}, and the scattering angle of a muon crossing the volume-of-interest (VOI) is obtained by using these two vectors as follows~\cite{carlisle2012multiple,nugent2017multiple,poulson2019application}
\begin{equation}
\theta=\arccos\left (\frac{\vec{v}_{1} \cdot \vec{v}_{2}}{\left|v_{1}\right|\left|v_{2}\right|}\right)
\end{equation}
The mean scattering angle due to the VOI and its standard deviation over $N$ number of the non-absorbed/non-decayed muons is calculated as written down in
\begin{equation}
\bar{\theta}\pm\delta\theta=\frac{1}{N}\sum_{i=1}^{N}\theta_{i}\pm\sqrt{\frac{1}{N}\sum_{j=1}^{N}(\theta_{j}-\bar{\theta})^{2}}
\end{equation}
Furthermore, the root-mean-square (RMS) of the scattering angle over $N$ number of the non-absorbed/non-decayed muons is deternined by using the following equation:
\begin{equation}
\theta_{\rm RMS}=\sqrt{\frac{1}{N}\sum_{i=1}^{N}\theta_{i}^{2}}
\end{equation}
On top of the scattering angle, we precisely register the number of the absorbed muons within the VOI as denoted in
\begin{equation}
\#_{\rm Capture}^{\rm In-target}=\mbox{\# of muMinusCaptureAtRest in VOI}
\end{equation}
Lastly, our simulation features are tabulated in Table~\ref{features}, and we conduct our GEANT4 simulations over a set of slabs including aluminum, copper, iron, lead, and uranium with the dimensions of 40$\times$10$\times$40 $\rm cm^{3}$ in accordance with another study~\cite{hohlmann2009geant4}. The particle tracking is managed by G4Step, and the registered hit locations on the plastic scintillators are post-processed by dint of a Python script.
\begin{table}[H]
\begin{center}
\caption{Simulation properties.}
\begin{tabular}{c c}
\toprule
\toprule
Particle & $\mbox{\textmu}^{-}$\\
Beam direction & Vertical\\
Momentum direction & (0, -1, 0)\\
Source geometry & Planar\\
Initial position (cm) & ([-0.5, 0.5], 85, [-0.5, 0.5])\\
Particle injector & G4ParticleGun\\
Number of particles & $10^{5}$\\
Energy distribution & Non-linear discrete\\
Target geometry & Rectangular prism\\
Target volume (cm$^{3}$) & 40$\times$10$\times$40\\ 
Material database & G4/NIST\\
Reference physics list & FTFP$\_$BERT\\
\bottomrule
\bottomrule
\end{tabular}
\label{features}
\end{center}
\end{table}
\section{Simulation results}
\label{results}
\begin{figure}[H]
\begin{center}
\includegraphics[width=7.9cm]{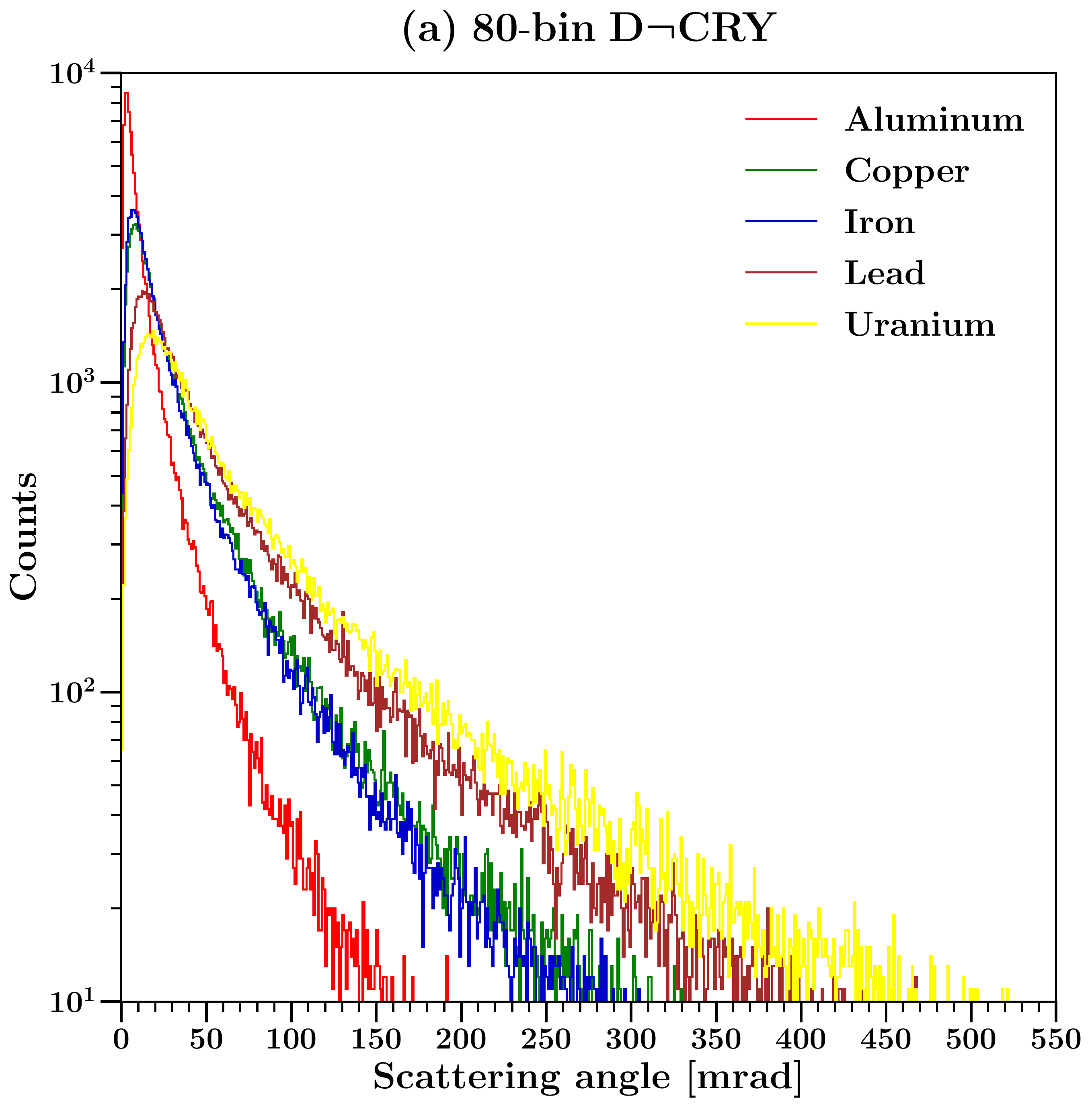}
\includegraphics[width=7.9cm]{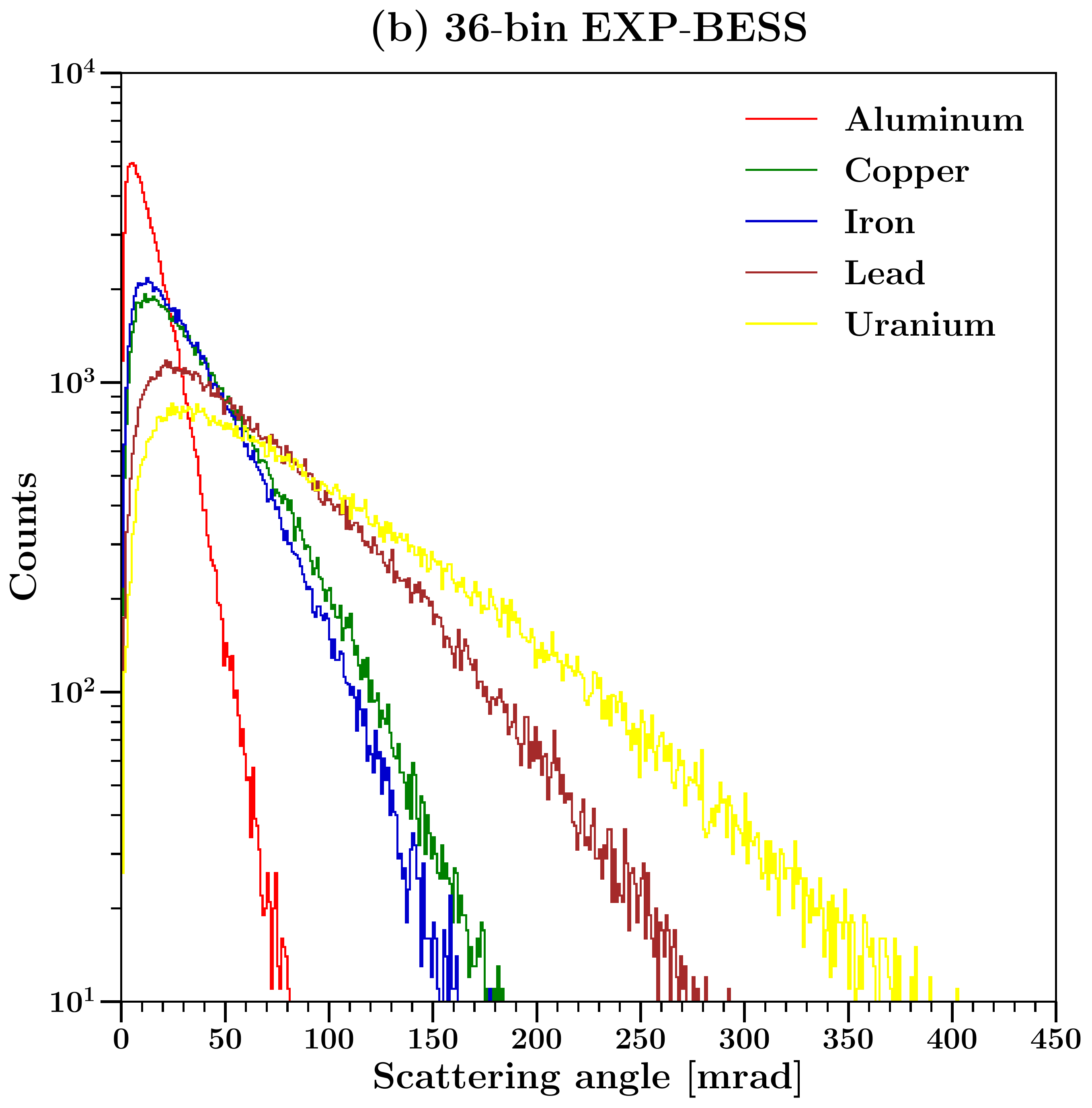}
\vspace{-2em}
\caption{Difference between the angular counts by using a step length of 1 mrad (a) 80-bin D$\neg$CRY and (b) 36-bin EXP-BESS.}
\label{anglecounts}
\end{center}
\end{figure}

In the long run, we inaugurate our GEANT4 simulations founded on our discrete approach, the initial outcomes of which are displayed in Fig.~\ref{anglecounts} in terms of the angular counts. From Fig.~\ref{anglecounts}(a)-(b), we explicitly see that the angular populations characteristically vary depending on the atomic number as well as the material density for the same thickness. Moreover, 80-bin D$\neg$CRY yields a typical exponential decrease towards the higher angles, whereas this trend appears to be a parabolic reduction in the case of 36-bin EXP-BESS.
Then, we also determine the average scattering angle as well as the RMS of the scattering angle in addition to the muon absorption and we compare both the discrete muon sources, i.e. 80-bin D$\neg$CRY and 36-bin EXP-BESS, as listed in Table~\ref{comparisoncharac}. From the contrast shown in Table~\ref{comparisoncharac}, we initially observe that the average scattering angle acquired via these discrete distributions is alike except in the case of lead and uranium. Secondly, 80-bin D$\neg$CRY constantly results in a strikingly higher standard deviation in comparison with 36-bin EXP-BESS since the first five energy bins of 80-bin D$\neg$CRY constitute a low-energy population excluded in 36-bin EXP-BESS that drastically amplifies the propagated uncertainty. Furthermore, this increased tendency is also examined in the RMS values owing to the same reason. Crucially, an essential fact is traced in the event of the muon absorption, and we demonstrate that 36-bin EXP-BESS does not yield any muon capture within the VOI since the energy level above 0.598 GeV has no potential for the muon absorption when the material thickness is merely 10 cm by recalling that the lower bound of 36-bin EXP-BESS is 0.598 GeV, while the lowest bin value in 80-bin D$\neg$CRY belongs to 0.1 GeV.
\begin{table}[H]
\begin{center}
\caption{Characteristic parameters obtained via 80-bin D$\neg$CRY as well as 36-bin EXP-BESS.}
\vskip -0.4cm
\resizebox{1\textwidth}{!}{\begin{tabular}{*7c}
\toprule
\toprule
Material & $\bar{\theta}_{\rm D\neg CRY}\pm\delta\theta$ [mrad] & $\theta^{\rm RMS}_{\rm D\neg CRY}$ [mrad] & $\#_{\rm D\neg CRY}^{\rm Capture}$ & $\bar{\theta}_{\rm EXP\mbox{-}BESS}\pm\delta\theta$ [mrad] & $\theta^{\rm RMS}_{\rm EXP\mbox{-}BESS}$ [mrad] & $\#_{\rm EXP\mbox{-}BESS}^{\rm Capture}$ \\
\midrule
Aluminum&15.980$\pm$27.004&31.378&-&15.036$\pm$12.413&19.499&- \\
Copper&40.638$\pm$60.312&72.725&1230&40.759$\pm$31.898&51.757&-\\
Iron&35.824$\pm$51.548 &62.773&1222&36.200$\pm$28.518&46.084&-\\
Lead&65.325$\pm$88.323&109.856&1272&68.138$\pm$52.450&85.987&-\\
Uranium&81.822$\pm$100.321&129.457&3836&95.721$\pm$75.376&121.836&-\\
\bottomrule
\bottomrule
\end{tabular}}
\label{comparisoncharac}
\end{center}
\end{table}
\section{Conclusion}
\label{Conclusion}
To sum up, we state a procedure based on the multi-group energy approximation where we favor the utilization of the binned energy values connected with the discrete probabilities by means of a probability grid. Consequently, we gain the capability to control as well as to adjust our energy spectra according to our computational goals apart from the noteworthy computation times.
\section*{References}
\bibliographystyle{elsarticle-num}
\bibliography{Discrete.bib}
\end{document}